\documentstyle[11pt,newpasp,twoside,epsf]{article}
\def\edcomment#1{\iffalse\marginpar{\raggedright\sl#1\/}\else\relax\fi}
\reversemarginpar

\begin{document}
\title{High Resolution Ultraviolet Quasar Absorption Line Spectroscopy
of $z \sim 1$ Galaxy Group}
\author{Jie Ding}
\affil{Department of Astronomy \& Astrophysics, The Pennsylvania
State University, 525 Davey Laboratory, University Park, PA 16802}

\begin{abstract}
We used ultraviolet spectra from HST/STIS ($R=30,000$), together with
optical spectra from Keck/HIRES ($R=45,000$), to study the three
MgII-selected absorption systems at $z=0.9254$, $0.9276$, and $0.9342$
toward the quasar PG~$1206+459$. A multi-phase gaseous structure, with
low-ionization components produced in small condensations and
high-ionization ones in diffuse clouds, is indicated in all three
systems. Each system is likely to represent a different galaxy with
absorption due to some combination of interstellar medium, coronal
gas, halo gas, and high-velocity clouds.  Even with the improved
sensitivity of HST/COS, we will only be able to obtain high-resolution
ultraviolet spectra of the brightest quasars in the sky.  A larger
telescope with ultraviolet coverage will enable quasar absorption line
studies of hundreds of galaxies, including a wide range of galaxy
types and environments at low and intermediate redshifts.
\end{abstract}

\section{Introduction}
The lightbeams from distant quasars pass through various galaxies and
probe physical properties of gaseous structures in the galaxies.
Therefore, studying absorption features in the background quasar
spectrum is a unique method for tracing the cosmic evolution of the
universe.  Strong MgII absorbers are almost always associated with
luminous galaxies ($> 0.05L^*$).  Thus, using this tool we can study
the predecessors of the giant spiral and elliptical galaxies that we
see in the nearby universe.

Three MgII systems (A, B, and C), at redshifts $z=0.9254$, $0.9276$,
and $0.9342$, are found along the line of sight toward the quasar
PG~$1206+459$. In a previous study, Churchill \& Charlton (1999) found
the three systems to be multi-phase absorbers with MgII clouds
embedded in extended, high-ionization gas that gives rise to CIV, NV,
and OVI. Their analysis was based upon the combination of
low-resolution data from HST/FOS and high-resolution data from
Keck/HIRES.  However, many important issues remained unresolved, such
as (1) whether CIV, NV and OVI arise in the same layer of gas and
whether their profiles are smooth or have sub-structure; (2) whether
the high-ionization phase ``envelops'' the low-ionization phase or
whether it is offset in velocity; (3) whether the majority of SiIV in
the $z=0.9276$ system arises in a single MgII cloud that is similar to
a Milky Way high-velocity cloud.

In May 2001, we obtained a stunning high-resolution ($R=30,000$)
HST/STIS spectrum, covering Lya and high-ionization transitions SiIV,
CIV, and NV, for the three systems. Through photoionization modeling
of the various chemical transitions in this spectrum and in the
earlier Keck/HIRES spectrum, the metallicities, abundance patterns,
and ionization states of absorbing gas clouds have been constrained
(for details on the modeling technique, see Ding et al. 2002). The
results are presented in the following sections along with the
physical interpretations of individual phases of gas.

\section{System~A at $z=0.9254$}

Six clouds, spread over $v \sim 200$~km/s in velocity space, are
needed to fit the MgII absorption. Ionization parameters are
constrained to be $-2.8 \la \log U \la -2.5$ and the metallicity is
required to be super-solar, if a solar abundance pattern is assumed.
These clouds, all having $\log N(\rm HI) < 16$, contribute negligibly
to the partial Lyman limit break. The dotted lines in Figure~1
represent the contribution from the MgII phase, for our best model.

A diffuse phase is required to fit the residuals in CIV, NV, and OVI,
unaccounted for by the MgII phase. Collisional ionization is ruled out as
a mechanism to produce the NV absorption, due to the narrow features
in the high-resolution spectrum. The ionization parameters of the
seven high-ionization clouds are constrained to be $-1.5 \la \log U
\la -0.6$.  A super-solar metallicity is also required in this phase,
unless N is enhanced. In Figure~1, solid lines represent the
contribution from this diffuse phase.

Neither the low-ionization phase or the diffuse phase could fully
account for the residuals to the red of the CIV profiles.  An
additional component, with $b (\rm C) \sim 6$~km/s, seems to be
necessary.  This component could be photoionized or collisionally
ionized. A near-solar metallicity is required in either case. The
photoionized component, with an intermediate ionization parameter
$\log U \simeq -2$, is represented by the dashed-dotted curves in
Figure~1.

The large kinematic spread of the MgII, with no dominant component,
the lack of a Lyman limit break, and the multiple-component structure
in NV suggest that this system does not represent a traditional
disk/corona structure.

\section{System~B at $z=0.9276$}

The strong, blended MgII profiles could be fit with five narrow
components, spread over $v \sim 200~$km/s in velocity space. The
ionization parameters of the low-ionization phase clouds are $-3.2 \la
\log U \la -2.5$ and the metallicity is constrained to be $\log Z
\simeq -0.1$, by the partial Lyman limit break detected in the FOS
spectrum.  The dotted lines in Figure~1 represent the contribution
from this phase.

The residuals in CIV, NV, and OVI indicate a highly ionized, diffuse
phase. A broad component, with $b (\rm N) \sim 50~$km/s, is needed to fit
the smooth NV absorption.  An offset, narrower one, with $b (\rm C) \sim
14~$km/s, is also needed to account for the blue part of CIV. An
ionization parameter of $\log U \simeq -1.6$ and a metallicity of
$\log Z \simeq -0.6$ are required for both clouds for a photoionization
model. The solid lines in Figure~1 show the two diffuse components.
Collisional ionization is ruled out in this phase due to the observed
ratio of CIII to CIV absorption.

An additional component, with $b (\rm Si) \sim 10$~km/s, is required to
produce the observed SiIV absorption unaccounted for by either the MgII
phase or the diffuse phase, in the cloud furthest to the red. This
component could be photoionized or collisionally ionized. The
collisional component, with a temperature $\log T\sim 4.8$, is
represented by the dashed-dotted curves in Figure~1.

The MgII phase is likely to arise in the disk of a galaxy. The broad
component in the diffuse phase may represent a corona structure
similar to that of the Milky Way. Two candidate absorbing galaxies
appear in a WIYN image of the quasar field at $\sim25$~kpc and
$\sim35$~kpc.
 
\section{System~C at $z=0.9342$}

A weak, unresolved MgII cloud, aligned with the bulk of the CIV
absorption in velocity space, is separated from System B, by $\sim
1,000~$km/s (see Figure~1).  If the MgII cloud does not give rise to
CIV, the ionization parameter of this phase is constrained to be $\log
U \la -3$ and the metallicity is $\log Z \simeq -0.6$. If, instead,
the majority of CIV is produced by the MgII clouds, the ionization
parameter is constrained to be $-2.2 \la \log U \la -2$ and the
metallicity is close to the solar value.

Regardless of the ionization state of the MgII phase, the strong
absorption in OVI (detected in the low-resolution HST/FOS spectrum)
requires an additional, highly ionized component. If this component is
photoionized, the ionization parameter would either be $\log U \simeq
-1.5$ or $\log U \ga -0.7$, depending upon whether CIV arises in the
same phase or not. Alternatively, if the diffuse component is
collisionally ionized, the temperature of this phase is constrained to
be $\log T \ga 5.4$.  Figure~1 shows a model in which CIV is produced
in the MgII phase and OVI (not shown) in a more highly photoionized
phase. The dotted lines represent the model contributions from the
MgII cloud, and the solid lines show those from the diffuse component.

In addition, a cloud offset $\sim -40~$km/s from the center of the
MgII absorption is required by the Lya profile. This cloud also gives
rise to the blue part of the CIV profiles. The dashed-dotted curves in
Figure~1 represent this offset cloud.

The simple, weak absorption feature suggests that this system may be a
high-velocity cloud or a dwarf galaxy. Two candidate associated
galaxies are found in the WIYN image. They would be dwarfs and have
impact parameters of $20$--$40$~kpc, if at $z\sim1$.

\begin{figure}
\plotone{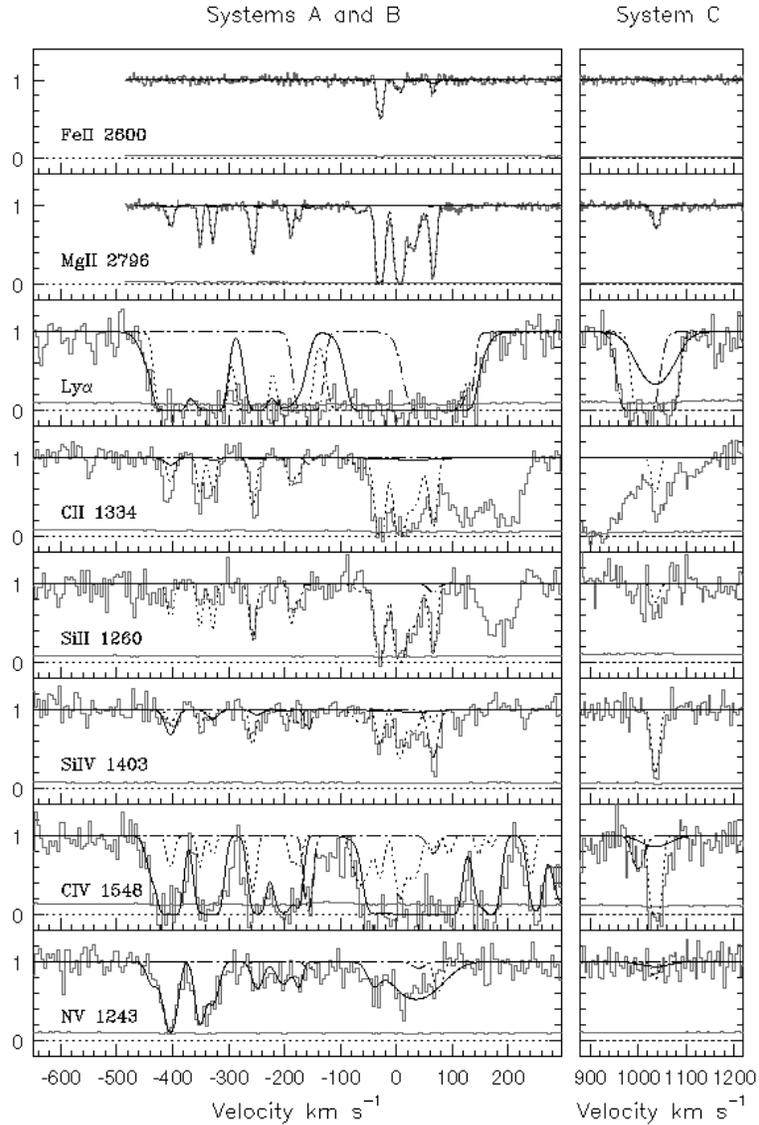} 
\caption
{Various key transitions are displayed in velocity space, with the
velocity zero-point at $z=0.9276$. The solid histograms represent the
normalized HST/STIS and Keck/HIRES data. The dotted lines show the
contribution from the low-ionization phase in each system. The solid
curves represent the photoionized, diffuse phase. The dashed-dotted
lines show the intermediate phase.}
\end{figure}

\end{document}